\documentclass[journal]{bugfixedIEEEtran}

\usepackage[pdftex]{graphicx}
\usepackage[utf8x]{inputenc}
\usepackage{url}
\usepackage{subcaption}
\usepackage{float}[width=\columnwidth]
\usepackage{siunitx}
\usepackage{tabularx}
\usepackage{amsmath}
\usepackage{afterpage}
\usepackage{balance}
\usepackage{amsmath}
\usepackage{amssymb}
\usepackage{multicol}
\usepackage{fancyhdr}
\usepackage[T1]{fontenc}

\renewcommand\IEEEkeywordsname{Index Terms}

\makeatletter
\newcommand{\setcaptype}[1]{\def\@captype{#1}}
\makeatother
\newsavebox{\tempbox}

\begin{document}

\title{\emph{Beyond the Metaverse:\\ XV (eXtended meta/uni/Verse)}}


\author{Steve Mann$^1$, 
Yu Yuan$^2$,
Tom Furness$^3$,
Joseph Paradiso$^4$, and
Thomas Coughlin$^5$
\\
1. U. Toronto;
2. President-Elect, IEEE Standards Association
3. Founder of U. Washington HITLab
\\
4. MIT Media Lab
5. IEEE President-Elect 2023 / IEEE President 2024 / President, Coughlin Associates, Inc
}

\date{}

\maketitle

\let\thefootnote\relax\footnotetext{\noindent Presented at IEEE Standards Association, Behind and Beyond the Metaverse: XV (eXtended meta/uni/Verse), Thurs. Dec. 8, 2022, 2:15-3:30pm, EST.}

\fancyfoot{
Presented at IEEE SA Behind and Beyond the Metaverse: XV (eXtended meta/uni/Verse), Thurs. Dec. 8, 2022.}

\begin{abstract}
We propose the term and concept XV (eXtended meta/omni/uni/Verse)
as an alternative to, and generalization of, the shared/social virtual reality widely known as ``metaverse''.
XV is shared/social XR.
We, and many others, use XR (eXtended Reality) as a broad umbrella term and concept to encompass all the other realities, where X is an ``anything'' variable, like in mathematics, to denote any reality,
X
∈
\{physical, virtual, augmented, \ldots \}
$\mathbb{R}$
eality.
%
Therefore XV inherits this generality from XR.
We begin with a very simple organized taxonomy of all these realities in terms of two simple building blocks: (1) physical reality (PR) as made of ``atoms'', and (2) virtual reality (VR) as made of ``bits''.
Next we introduce
XV as combining all these realities with extended society as a three-dimensional space and taxonomy of
(1) ``atoms'' (physical reality), (2) ``bits'' (virtuality), and (3) ``genes'' (sociality).
Thus those working in the liminal space between Virtual Reality (VR),
Augmented Reality (AR), metaverse, and their various extensions, can describe their work and research as existing in the new field of XV.
XV includes the metaverse along with
extensions of reality itself like shared seeing in the infrared, ultraviolet,
and shared seeing of electromagnetic radio waves, sound waves,
and electric currents in motors. 
For example, workers in a mechanical room can look at a pump and see a superimposed time-varying waveform of the actual rotating magnetic field inside its motor, in real time, while sharing this vision across multiple sites.
\end{abstract}


\section{Intro, history, and background}
The metaverse is shared/collaborative/social VR (Virtual Reality).
We propose XV as shared/collaborative/social XR (eXtended Reality).
XR is the overarching term and concept to:
{\bf (1) interpolate between, and encompass, all the other realities} including physical reality.
As in mathematics, let X be a variable to denote any reality,
X
∈
$\mathbb{R}$
eality,
∀ (for all) of them, PR (Physical Reality) VR (Virtual Reality), AR (Augmented Reality), etc., so we can eliminate or step beyond the jargon and confusion especially for those of us who work in the liminal space between and beyond all these realities,
and {\bf (2) eXtrapolate or eXtend beyond them}. Thus XR may be variously called ``eXtended Reality'', ``Cross Reality'', etc.\cite{mannwyckoff91,  paradiso2009guest, lifton2009metaphor, mann2018all, ratcliffe2021extended, mayton2017networked, mann2022swim}.

\subsection{Bringing order and structure to the $\mathbb{R}$ealities}
\begin{figure}
    \centering
    \includegraphics[width=\columnwidth]{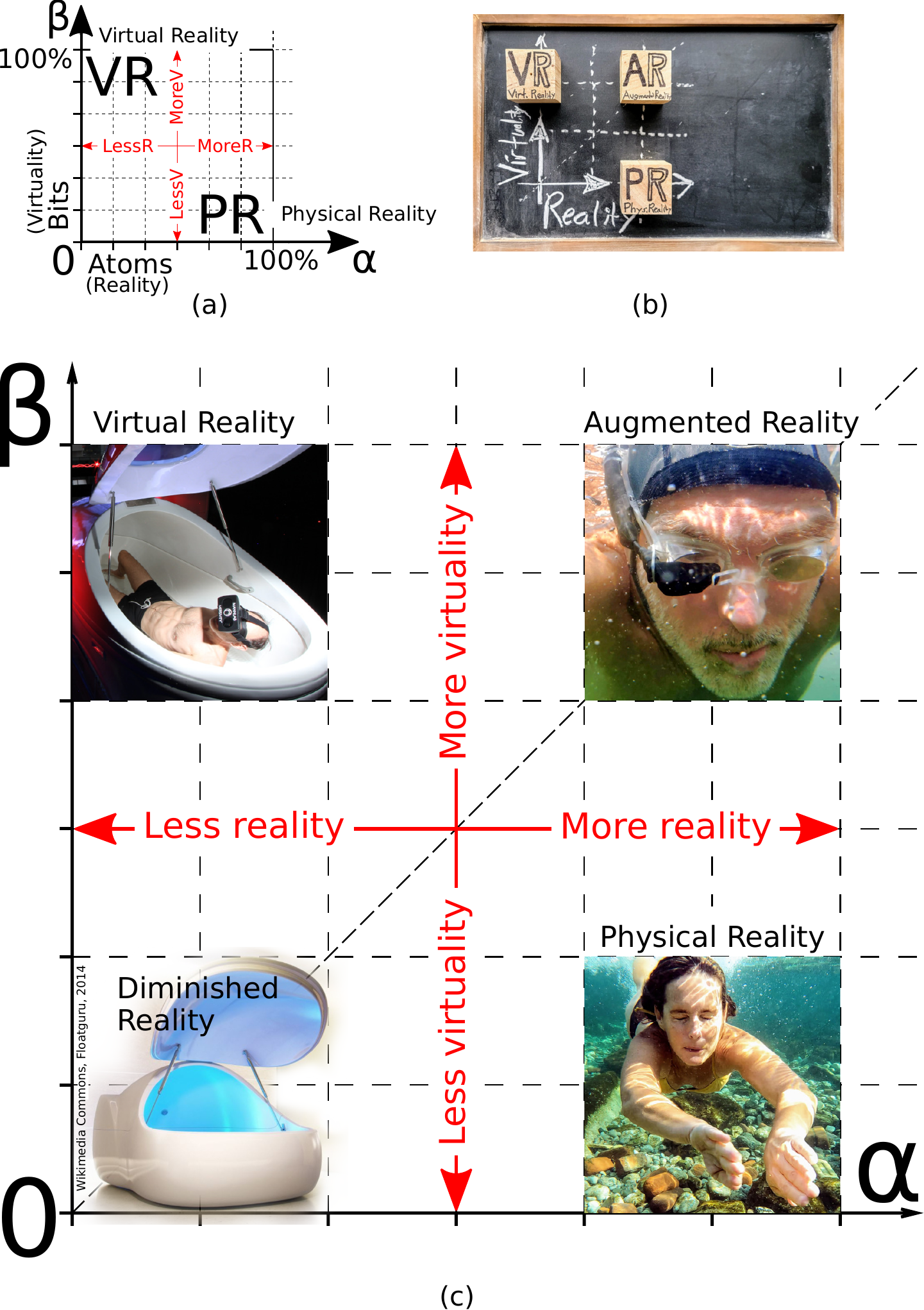}
    \caption{``{\bf The Realities}'' (a) Physical Reality (PR) exists in the physical world of ``Atoms''.  The word ``atom'' is of Greek origin, and begins with the letter $\alpha$, so we label the ``Reality'' axis ``$\alpha$''.  Virtual Reality (VR) exists in the virtual world of ``bits'' (in the Claude Shannon sense, i.e. units of digital {\em or analog} information), so we label the ``Virtuality'' axis ``$\beta$''. (b) Augmented Reality (AR) includes both Reality and Virtuality.  (c) This suggests a simple taxonomy for the realities.  Notice the lower left quadrant needs a name which we call Diminished Reality (DR).  Examples of DR include dark sunglasses, ear plugs, baseball caps for humans or blinders on horses, and of course sensory isolation tanks (``float tanks'').}
    \label{fig:axes}
\end{figure}

The term ``Virtual Reality'' (VR) was introduced in 1938 by Artaud~\cite{artaud} as an alternative reality we can think of as orthogonal to ``Physical Reality'' (PR), as illustrated in Fig~\ref{fig:axes}(a)~\cite{mann2022swim}.
Physical reality exists in the world made up of atoms.  The word ``atom'' is of Greek origin.
More than 2400 years ago, Greek philosopher Democritus proposed that all matter was made of discrete small particles called
$\alpha \tau o \mu o \varsigma$,
(``atomos'').
This word begins with the Greek letter $\alpha$ (alpha), so we say that reality exists along an
axis labelled ``$\alpha$''.
We recognize that the real-world universe includes not just objects like atoms that have mass, but also that which has energy, such as photons, and that the matter that makes up the mass of the universe is made of atoms which themselves are made of sub-atomic particles.
So we use the word ``atom'' in the traditional Greek sense.

VR exists in the world made of ``bits''.  We use the word ``bit'' in the wide (broad) sense, as suggested by Claude Shannon: ``bits = ``binary digits'' of information, whether that information is digital or analog
(the word ``bit'' in this sense was suggested by J. W. Tukey in 1947\cite{shannon1948mathematical}).
Thus the ``Bits'' ($\beta$) axis of Fig~\ref{fig:axes} represents the virtual/informatic world regardless of whether it be analog or digital.  Early work on virtual worlds, virtual hand controllers, etc., was at the liminal space between digital and analog~\cite{furness1986super}.
The resurgence in popularity of analog synthesizers~\cite{paradiso2017modular} also results in the kind of mashups between digital and analog technology that makes it necessary to broaden the concept of virtuality beyond only that which is digital (quantized), as well as beyond that which is necessarily visual
(e.g. we can have sound-based realties~\cite{mayton2017networked} and olfactory-based realities~\cite{zhao2020methods}).



Augmented Reality (AR), as pioneered by Ivan Sutherland~\cite{sutherland68}, and many others, exists in the real and virtual world, as illustrated in Fig~\ref{fig:axes}(b).
Mixed reality has also been proposed~\cite{milgram94, milgram99} as a one-dimensional taxonomy that slides continuously between VR and PR~\cite{zuo2020novel, tepper2017mixed},
but has no origin (no zero value).
When abbreviated it can be conflated with MR = Mediated Reality which is itself another important member of the realities~\cite{mann260, grasset2005interactive, ibrahim2019computer, poelman2012if, barbatsis1999performance, seeingWithTheBrain_barfieldMann, haller2005loose, tang2002seeing, barfieldbook7, hillfungmannicip, mann2018all}.

When we organize the Physical, Virtual, and Augmented Realities (PR, VR, and AR) in the two-dimensional $\alpha$-$\beta$ space, we generate a taxonomy divided into four quadrants, as in Fig~\ref{fig:axes}(b) with only 3 building blocks, and a missing block at the origin.
The missing lower-left quadrant could be filled with many examples.
There are a wide-range of technologies that are designed to deliberately diminish/reduce reality/realities.
Examples include earplugs, dark sunglasses and welding helmets, baseball caps to shield from bright overhead light, blinders on horses, and even cellphone jammers/shields.  These technologies belong in the lower left quadrant, near the origin, under a new category we call ``Diminished Reality'' (DR), as shown in Fig~\ref{fig:axes}(c).
Here we selected four examples that each illustrate a corresponding quadrant.
A sensory isolation tank (``float tank'') is used to represent the lower left ``Diminished Reality'' quadrant.  The experience of icewater swimming is used to represent the lower-right quadrant, ``PR'' (``Physical Reality''), since this experience is fully immersive and very difficult to mimic using VR.  The idea of wearing a VR headset in a sensory isolation tank has also been explored, as a way to remove a great deal of input from physical reality and thus make the immersive VR experience more purely along the $\beta$ axis~\cite{deconference2003, mann2022swim}.
Thus we use the VR float tank to represent the upper left quadrant.
To represent the upper right quadrant,
we use the example of icewater swimming with an underwater augmented reality headset.  This is a nice example of AR, as it involves a great deal of physical real-world stimulus combined with informatic content such as maps, wayfinding, warnings, water-temperature indication, brain signals, heart, etc., as augmented overlays.
This provides the swimmer with situational awareness of the surroundings as well as internal body state~\cite{mann2022water}.
Note that there are a very large number of different technologies that fit into each of the four quadrants, as well as many that fit in the liminal spaces between them, and using this simple taxonomy helps us sort through the massively confusing assortment of different kinds of realities.

Note also that this taxonomy does not directly address human perception.  For example, in an isolation tank, or more generally when we are presented with zero stimulus, we may enter into a hallucinogenic or lucid-dreamlike state and start to imagine things that are not really present.  Perception doesn't necessarily go to zero when stimulus does.  In this sense, our taxonomy is a taxonomy of technology in terms of ``what it is'' rather than ``what it does''.
Consider by way of example the concept of temperature.  Each of us perceives temperature differently.
For one person, a swim in a 4-degree Celcius lake might feel really cold, but to an icewater swimmer it might feel more normal.  The same person might even perceive the same temperature differently on different days, or even on a subsequent swim on the same day.
But temperature is still a useful concept, and, as a concept, allows us to separate the physics from perception so that we can make sense of the world in simple ways, while acknowledging that other more complicated models are required for full human conceptualization.  In this way our taxonomy is a very simple and elementary foundation upon which we hope that others will build.
Indeed, one of our Grand Challenges is human-factors and perception, which we will expand upon shortly.

Moreover, there are other dimensions beyond $\alpha$ and $\beta$, which we will address shortly.

\begin{figure}
    \centering
    \includegraphics[width=0.75\columnwidth]{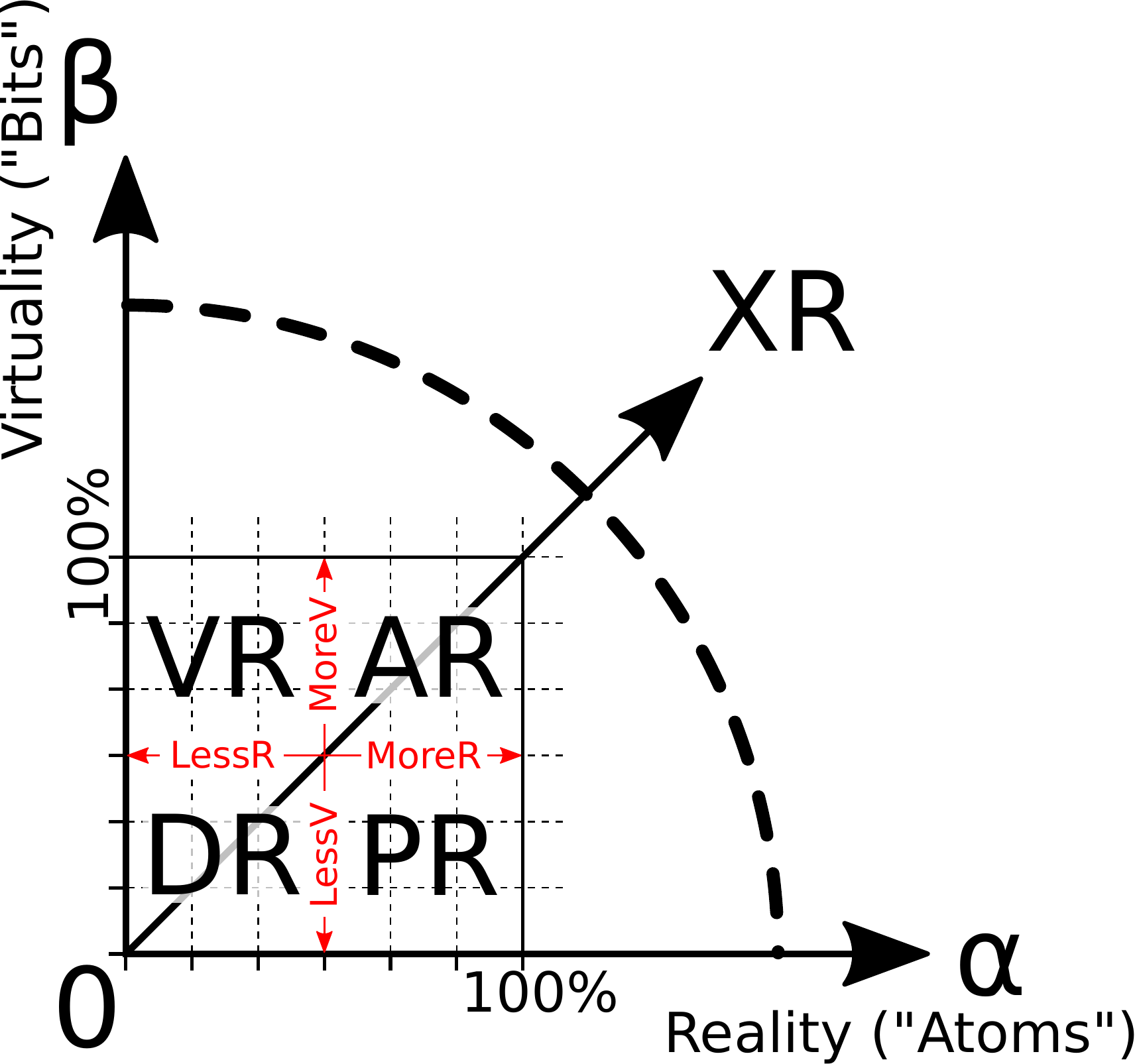}
    \caption{XR = eXtended Reality aims to {\bf (1) interpolate between, and subsume}, existing ``realities'' (PR = Physical Reality, VR = Virtual Reality, AR = Augmented Reality, DR = Diminished Reality, etc.), and {\bf (2) extrapolate} beyond them.}
    \label{fig:XR}
\end{figure}

Even within the $\alpha$-$\beta$ plane, we can think beyond these four quadrants!  That's an important goal of XR (eXtended Reality): to extend the boundaries of the $\alpha$-$\beta$ plane.  See Fig.~\ref{fig:XR}.

\subsection{XR: interpolating and eXtrapolating}
The goal of XR is twofold, specifically so that we can:
\begin{enumerate}
    \item {\bf interpolate} between and subsume all the existing realities such that we can generalize,
e.g. so that researchers working at the liminal spaces between the realities can use the term ``XR'' as a broad ``catch-all'' term and say something like ``I work in the field of XR.''; and
    \item {\bf eXtrapolate} beyond the existing realities~\cite{mann2022swim}.
\end{enumerate}
These goals are illustrated in Fig~\ref{fig:XR}.
Moreover, XR can be (and often is) used to interpolate along one axis and extrapolate along the other.
A good example is HDR (High Dynamic Range) imaging which allows us to see beyond the range that a normal human can see, while also providing a very small amount of virtual content such as not to distract the user from reality.  HDR is considered a quintessential example of XR\cite{mannwyckoff91} where the reality is extended along the $\alpha$ axis beyond 100\% but often with only a small or even 0\% virtual content ($\beta = 0).$

Another example is icewater swimming while wearing an underwater augmented reality head-up display.
This gives us a heightened, rather than diminished, capacity to engage in icewater swimming, but we often program it to maintain a lesser virtual presence such as to put the main emphasis on reality and safety.

This way of using XR (eXtended Reality) to get more than 100\% reality is possible because an awareness of our physiological conditions helps us be even more ``in the moment'' while calming our breathing, reducing
our heart rate using biofeedback, and enhancing our awareness of our surroundings using minimalist but important overlays.

\begin{figure}
    \centering
    \includegraphics[width=\columnwidth]{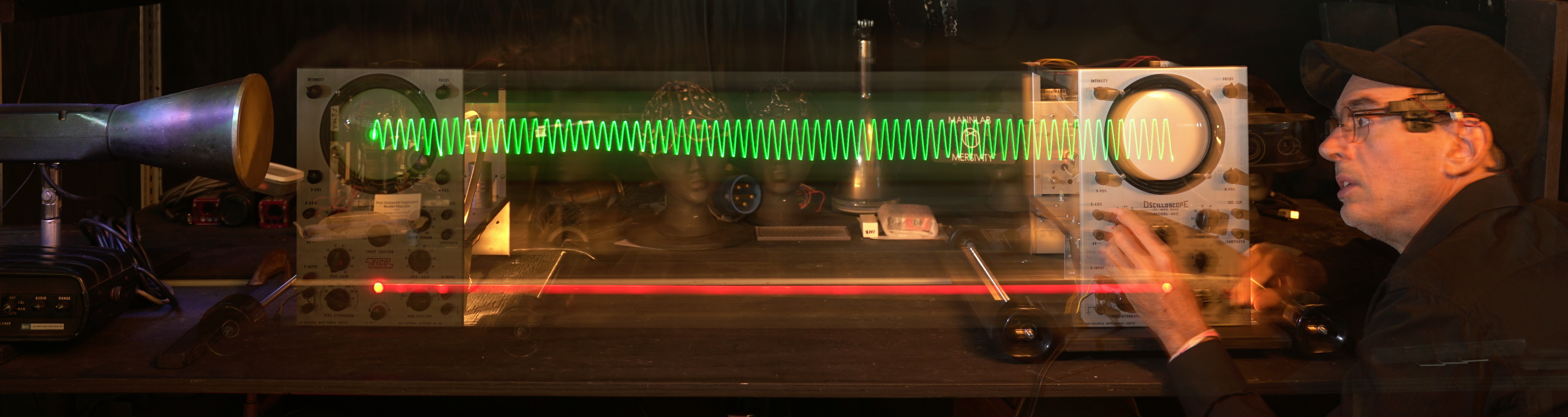}
    \caption{Discovery of the Metavision / Metaveillance principle: Photograph of cathode-ray tube (in an oscilloscope) moving in front of a police radar, while connected to the baseband Doppler output of the radar.}
    \label{fig:kustom}
\end{figure}

Another example of XR is SWIM (Sequential Wave Imprinting Machine, as shown in Fig.~\ref{fig:kustom})
which extends human perception 
into the ultraviolet, infrared, ultrasonic, infrasonic, etc., and allows us to see radio waves, sound waves, and, perhaps more profoundly, see or sense sight or sensing itself (metavision).
This allows us to inspect smart self-driving cars to see if their sensory systems are in good order,
or to observe the performance of water pumps in a mechanical room by looking into the rotating magnetic fields in the pump motors.

\subsection{Metavision, Metaveillance, and the Metaverse}
The SWIM (Sequential Wave Imprinting Machine) was invented in Canada in 1974 as a form of XR (eXtended Reality) and Metavision / Metaveillance~\cite{mannwyckoff91, pereira2019extended, ccoltekin2020extended, chuah2018and, kenwright2020future}.
Metavision is
the realtime display or photography of electromagnetic waves (e.g. radio waves as shown in Fig.~\ref{fig:kustom}),
sound waves, and the like~\cite{swim2018}, as well as
the realtime display or photography of the electrical signals associated with rotating magnetic fields in electric machines~\cite{mann2020moveillance, mann2020visualizing}, as well as the realtime display or photography of sound waves and other propagatory waves in solid, liquid, gas, or other matter.  SWIM has origins in marine radar and sonar (e.g. display or photography of sound waves underwater), and was also an early example in the new field of WaterHCI (Water-Human-Computer Interaction)~\cite{mann2021water, mann2022water}.

Most interestingly, SWIM went beyond being a means to see and photograph waves, but also was a kind of meta-sensing device, i.e. a system to sense sensors and sense their capacity to sense.

The word ``meta'' is a Greek word that means ``beyond'', in a self-referrential sense, e.g. a meta-conversation is a conversation about conversations.  A meta argument is an argument about arguments.  Metadata is data about data (e.g. the header of an image file that indicates where the picture was taken, shutter speed of the camera, etc.).

Meta-sensing is the sensing of sensors and the sensing of their capacity to sense.
For example, a radar is a sensor that police use for surveillance.
A radar detector is a device that senses radar sensors, and
a radar detector detector is a device that police use to sense if
someone is sensing them doing the sensing.

Metavision is a way of visualizing the capacity of a sensor to sense.
The concept of metavision was discovered in 1974 when the Doppler signal of a radar was connected to an oscilloscope that had no timebase, and the oscilloscope was moved back-and-forth in front of the radar, causing a Doppler shift in the signal being displayed.
This close feedback between human and machine resulted in an immediate display that extended human vision to allow humans to see and photograph radio waves at the speed of light.
See Fig~\ref{fig:kustom}.

SWIM can be digital or analog, and in fact was originally completely analog (not digital).  Thus it serves as a useful example of eXtended Reality XR in contrast to the virtual and ``Digital Reality\cite{mcdermott2012copyright}'' of the metaverse~\cite{snowcrash,dionisio20133d}.

\begin{figure}
    \centering
        \centering
        \includegraphics[width=\columnwidth]{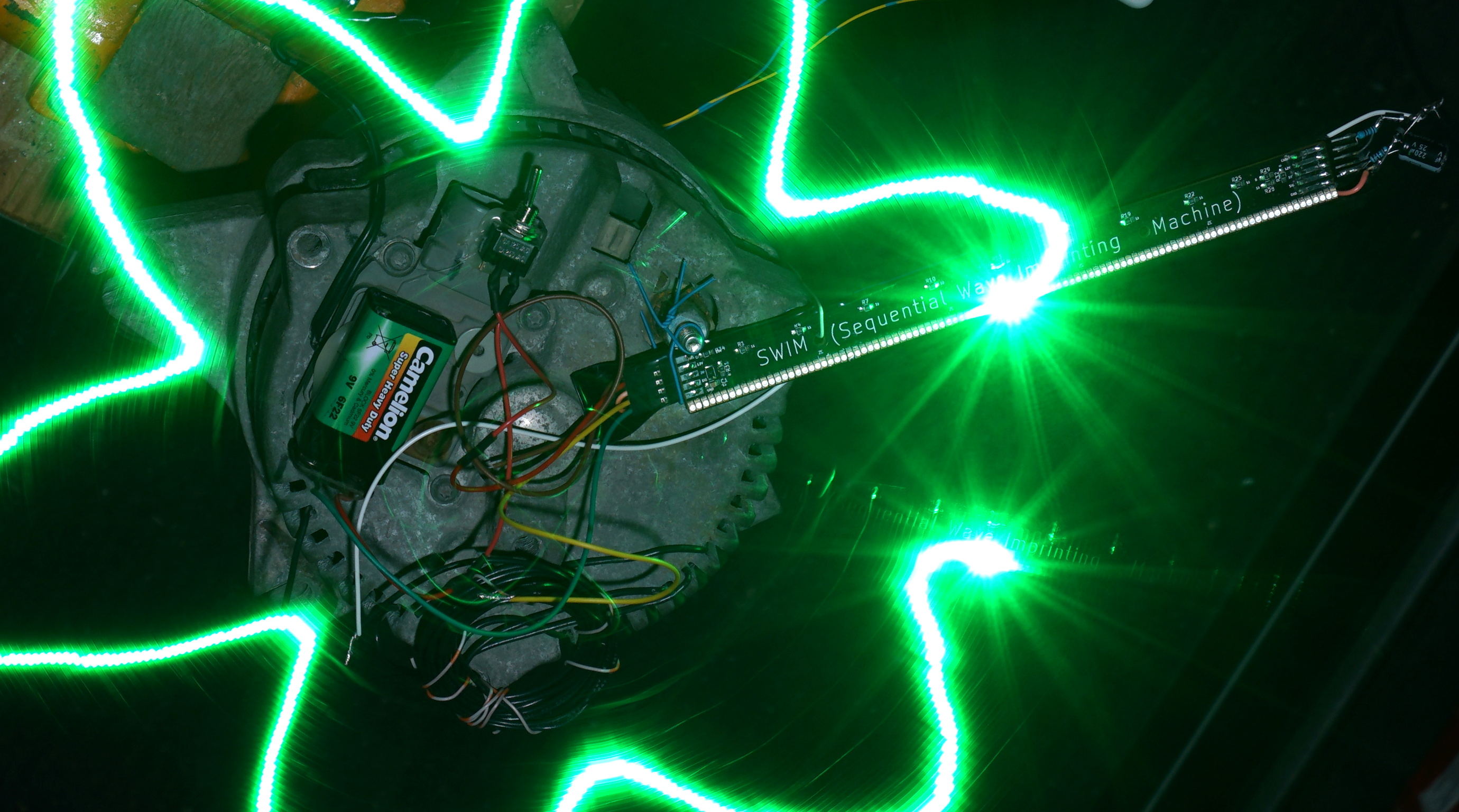}
    \caption{Rotary SWIM allows users to safely see and interact with voltages or currents associated with rotating magnetic fields~\protect\cite{mann2020visualizing}.}
    \label{fig:swimotor}
\end{figure}

The example shown in Fig~\ref{fig:kustom}
is an example of {\em linear} SWIM, whereas other examples of SWIM also include rotary SWIM.
Previous work in making the inner workings of electric machines visible
includes ``Moveillance''\cite{mann2020moveillance, mann2020visualizing}
which reveals in real-time the inner workings of a motor, using rotary SWIM (one or more linear arrays of LEDs mounted
radially outwards from the motor).
See Fig.~\ref{fig:swimotor}.
The result is a polar oscilloscope of sorts that is perfectly aligned with the physical body of the motor.

\section{XV happens when the 3rd axis is added}
\begin{figure*}
    \centering
    \includegraphics[width=\textwidth]{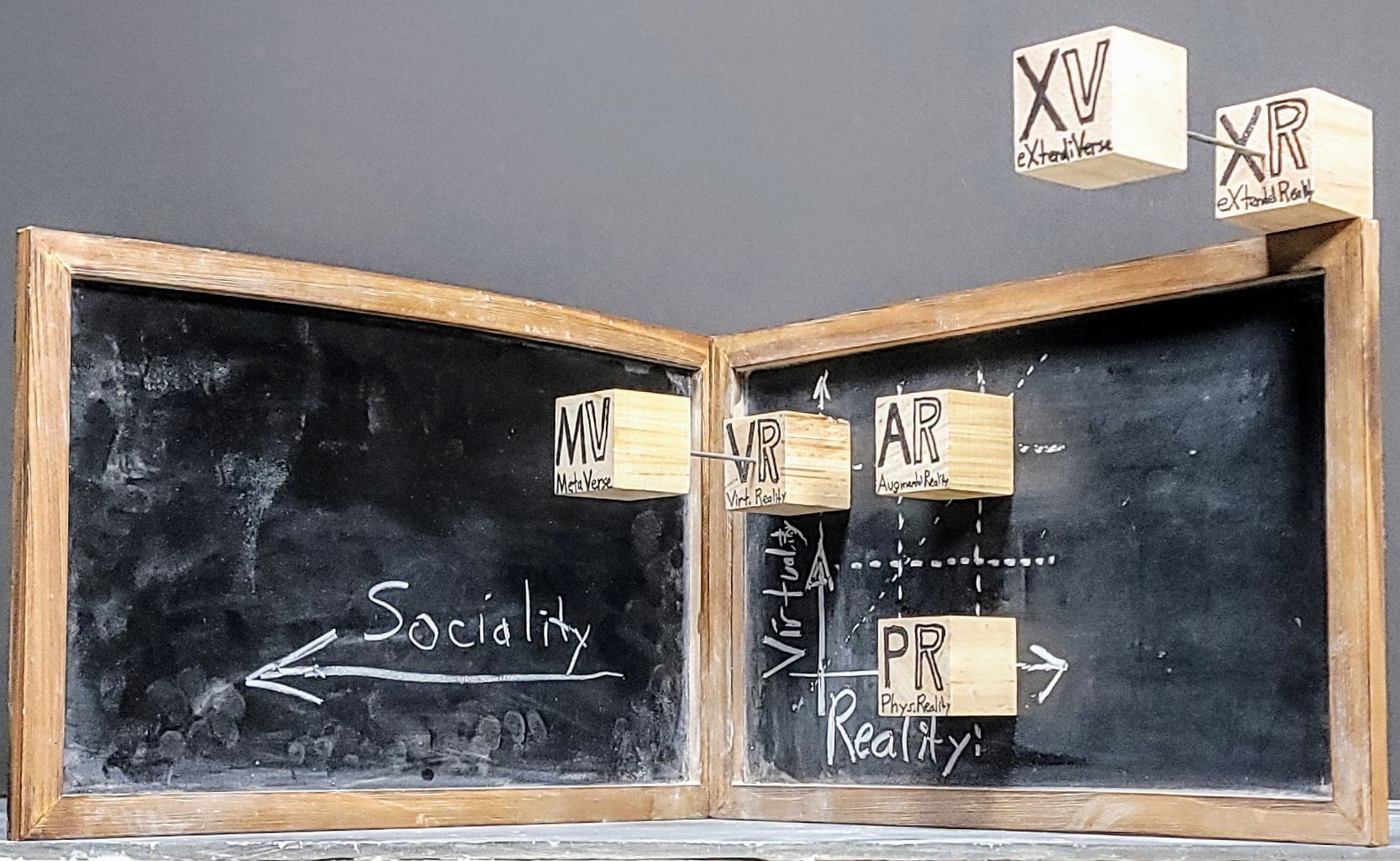}
    \caption{
     {\bf The building blocks of the XV continuum.}
     The three axes of the continuum are: (1) Reality (atoms, $\alpha$); (2) Virtuality (bits, $\beta$); and (3) Sociality (genes, $\gamma$).
     The four blocks (PR, VR, AR, and XR) in the
     Reality-Virtuality plane are in the plane of the two axes drawn on the chalkboard.  Compare with Fig.~\protect\ref{fig:axes}(b).
     The metaverse extends out of this plane, along the third (social) dimension, from VR, i.e. the metaverse is shared/social/collaborative VR.
     The eXtendiVerse (XV) extends out of the Reality-Virtuality plane from XR, i.e. XV is shared/social/collaborative XR.
    }
    \label{fig:blocks3d}
\end{figure*}
Now that we understand XR and some examples of XR that go beyond the other realities, let us now add a third axis (dimension) to the $\alpha$-$\beta$ plane, turning it into a 3-dimensional volume (volumetric taxonomy rather than just a planar taxonomy), as shown in Fig.~\ref{fig:blocks3d}.

The third axis is human sociality.  Genes are the smallest unit of humanness, and ``gene'' is a word of Greek origin, 
from Ancient Greek $\gamma \epsilon \nu \epsilon \alpha$.
The first letter of this word is the Greek letter $\gamma$ (gamma),
which is the third letter of the Greek alphabet.

Thus we now have the three axes, atoms, bits, and genes,
denoted by the first three letters of the Greek alphabet,
$\alpha$, $\beta$, and $\gamma$.

The third axis denotes a human societal axis, the human element of emotional support and togetherness, as shown in Fig.~\ref{fig:blocks3d}.
This goes beyond our earlier analogies of the $\alpha$-$\beta$ plane as the Argand plane, taking us instead into the kind of 3-dimensional taxonomy that can be easily visualized using the very technologies that it is categorizing -- very ``meta'' (self-referrential) indeed.

\begin{figure}
    \centering
    \includegraphics[width=\columnwidth]{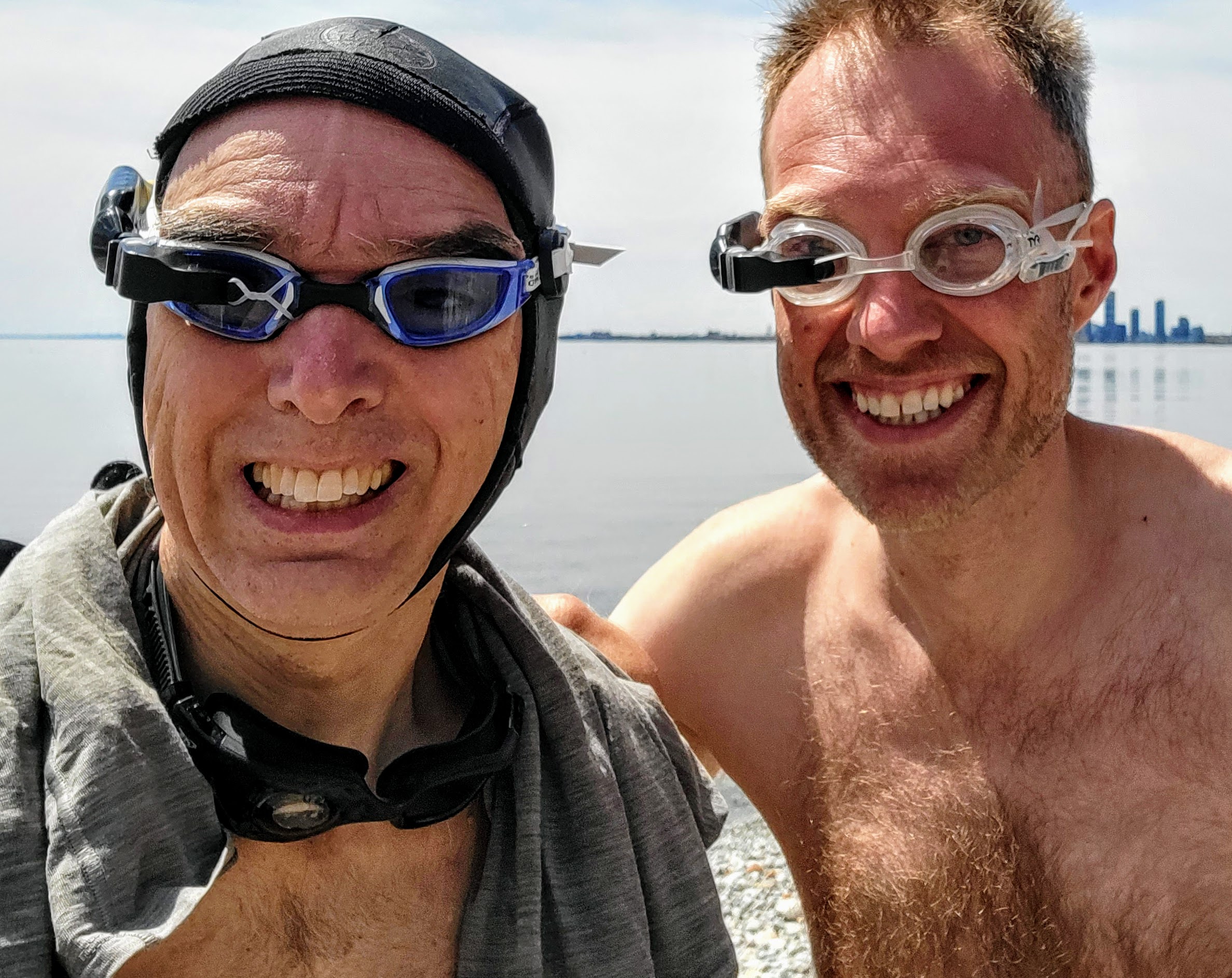}
    \caption{
    Example of XV (shared social collaborative XR):
    doing icewater swimming together as a group, wearing immersive XR (eXtended Reality) eyewear to communicate safety-related information, and share maps, overlays, safety messages, and warnings, etc.}
    \label{fig:smartswims}
\end{figure}

\begin{figure*}
    \centering
    \includegraphics[width=\textwidth]{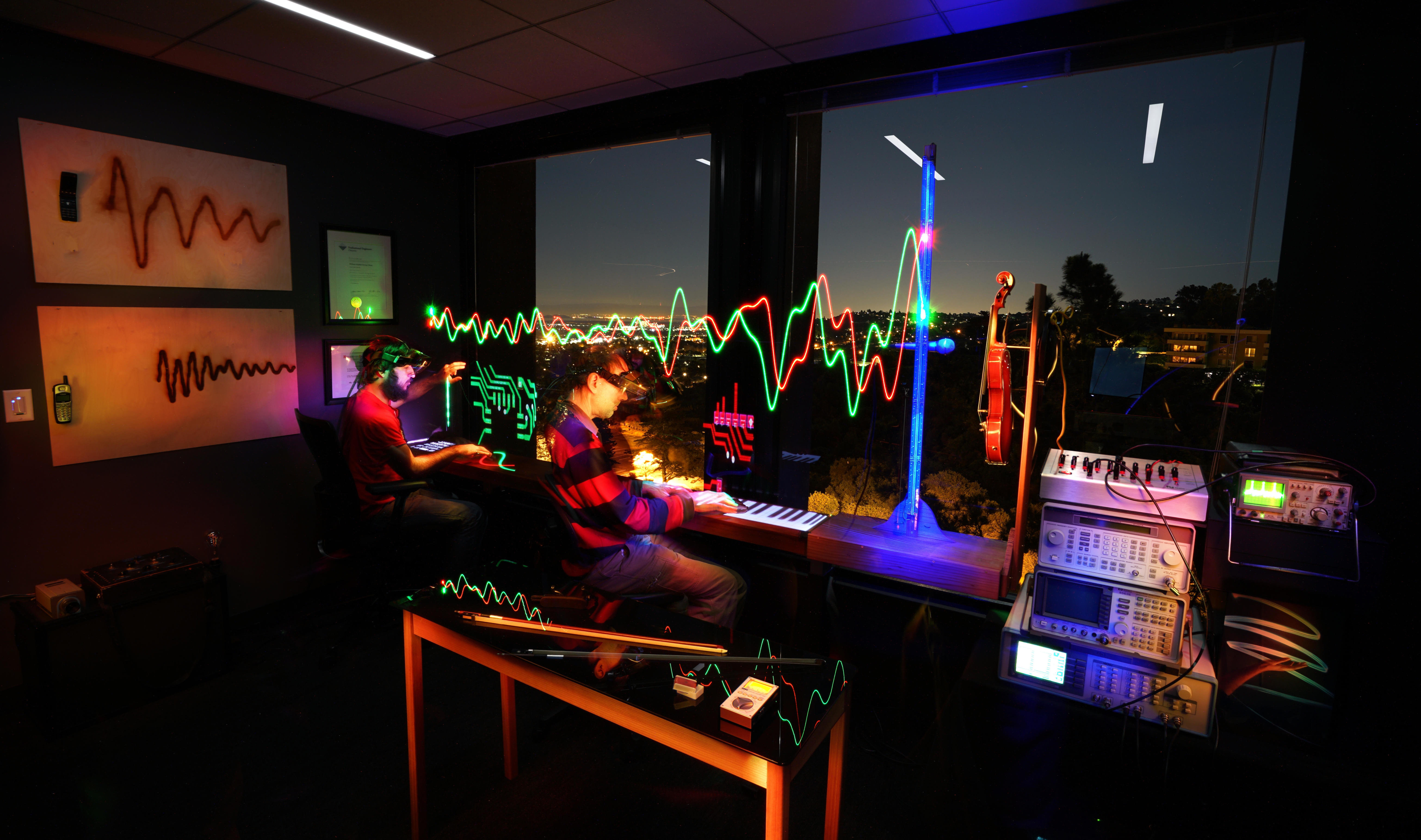}
    \caption{
    Examples of XV (shared social collaborative XR):
    working together on the design of a violin while looking at the sound waves coming from it, using SWIM (Sequential Wave Imprinting Machine),
    as well as designing some sound processors and associated circuit boards.
    This picture of Steve Mann in his office at Meta, collaborating with Alex Papanicolaou (both wearing the {\em Meta 2} eyewear), was taken in 2017 by Mann and Rob Godshaw working together to capture the SWIM which Alex helped to build.
    }
    \label{fig:alexsteve}
\end{figure*}

Some examples of shared/social/collaborative XV are shown in Fig.~\ref{fig:smartswims}
and Fig.~\ref{fig:alexsteve}.

The SWIM pictured in the the latter example (Fig~\ref{fig:alexsteve})
allows others in the room who are not wearing an eyeglass (in this case, the {\em Meta 2} that both people are wearing to collaborate) to also see much (but not all) of what is happening in XV.
In this way, those wearing the technology as well as those without it can
collaborate at multiple scales both locally (in the room) and remotely (at other geographical locations, even in other countries).  In this way XV operates at multiple scales from microscopic to worldwide~\cite{mann2020sensing}.

In regards to social scale and distance along the social axis,
we have also explored the world of performance art across various scales of space and time, as, for example, in the Social Distancer~\cite{arxiv2021vironment} and Equinox~\cite{ramsay2022equinox}.

\subsection{Scale-based taxonomies}
\begin{figure}
    \centering
    \includegraphics[width=\columnwidth]{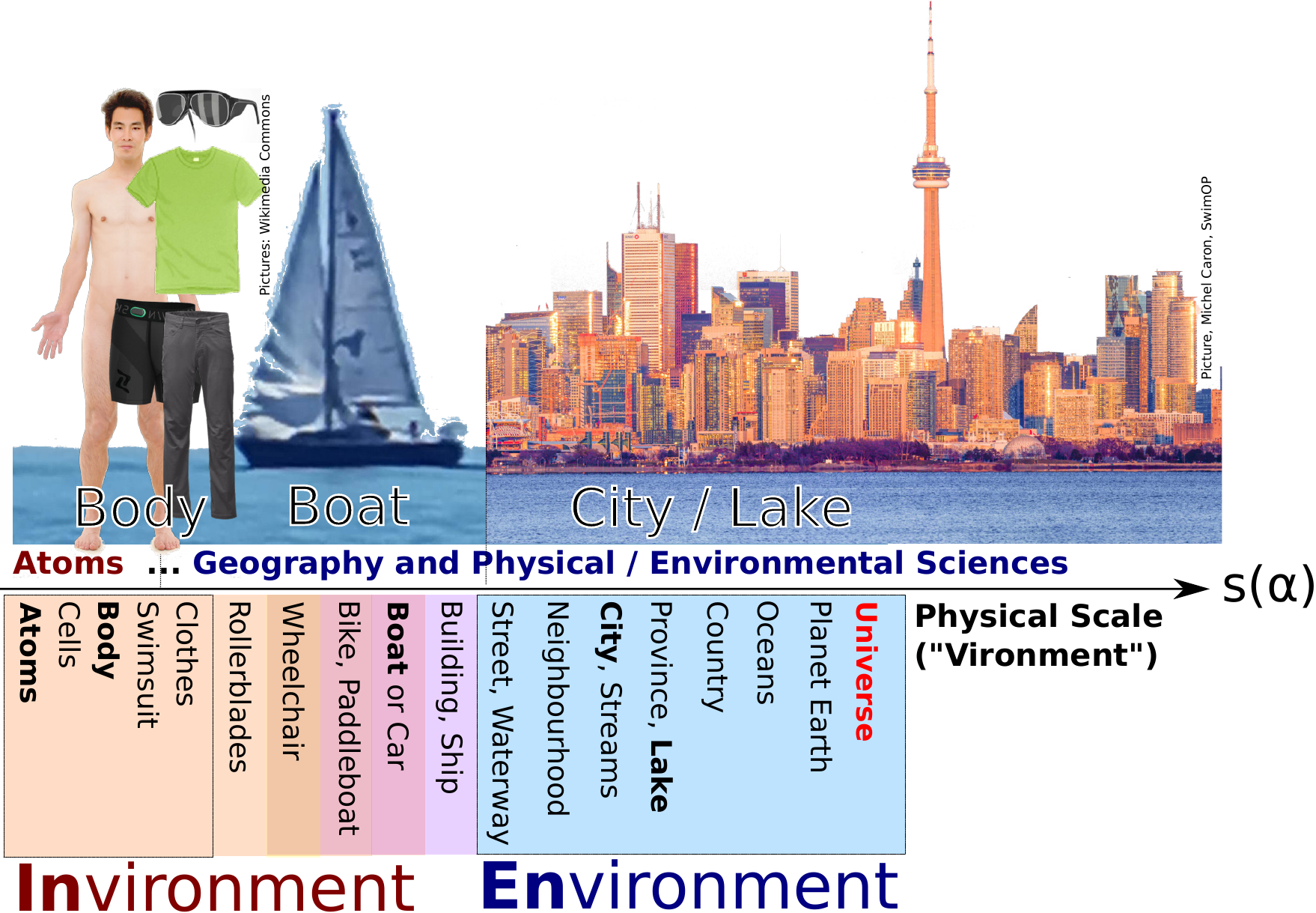}
    \caption{Introducing the concept of scalespace: The Reality ($\alpha$) axis now becomes the Physical Scale of reality axis, $s(\alpha)$.  Atoms are close to the origin on the $s(\alpha)$ axis.  The edge of the Universe is further out on the $s(\alpha)$ axis.
    A natural taxonomy results that separates wearables (wearable technologies, i.e. technologies of our Invironment = technologies that we consider to be part of us) from technologies that surround us (i.e. technologies of our Environment, i.e. technologies we usually don't consider to be part of us).
    The {\em Environment} is denoted as blue and the {\em Invironment} is denoted as fleshtones.
    }
    \label{fig:scalespace_vironment_only}
\end{figure}

So far the XV taxonomy has been presented as a three-dimensional space in which the three axes ``atoms'', ``bits'', and ``genes'' represent {\em quantities}.
Now let us consider the taxonomy in terms of scale, i.e. how big or small something is.
The Reality axis shall now represent the physical scale ranging from atoms near the origin,
to the edge of the universe, as shown in Fig.~\ref{fig:scalespace_vironment_only}.
This provides a natural taxonomy with two categories: technologies that are part of us,
and technologies that are part of our environment (surroundings).
The relatively sharp split between the small scales (Invironment) and the large scales (Environment) is nicely captured by some simple everyday examples as well as an example from popular culture:
\begin{itemize}
 \item In the local marina or yacht club (or in a parking lot) we often hear boaters (or motorists) say ``You hit me!'' when their vessels (or cars) collide or allide (personal observations -- within the field of W.H.A.T. = Wearables, Humans, And Things, as jointly proposed by N. Gershon and S. Mann~\cite{mann2015what}).
 Indeed, Manfred Clynes, who coined the term ``cyborg'' (cybernetic organism) said that his favorite example of a cyborg is a person riding a bicycle, and we note the same ``You hit me!'' effect with bicycles.
 It has also been suggested that boaters are also cyborgs and thus cyborgs existed more than a million years ago!~\cite{mann2021water}  The word ``cybernetic'' is in fact a word of Greek origin that means ``helmsman''.
 \item In science fiction, when people spontaneously materialize across the spacetime continuum, there is an implicit default assumption that their clothes, but not their surroundings, will travel with them, so much so, that it would seem unusual -- humourous even -- for technologies of the Invironment, like clothes, to not travel with their wearer:
 ``Very funny, Scotty. Now beam me up my clothes.''~\cite{scottyclothes}
\end{itemize}

Now consider also the informatic scale (Virtuality) ranging from ``bits'' near the origin, through ``little data'' like distributed blockchain and extending out to ``big data'' like centralized databases as with a central bank).

Now consider the scale of the sociality axis, ranging from Genes (smallest unit of humanness),
to individuals, small communities, etc., to larger and larger social structures.
In regards to social scale, the concept of ownership is most prevalent, e.g. ownership of data such as copyright, as might be associated with photographs or video recordings.
We have thus a dichotomy of social ownership scale in the division between
surveillance (``big watching''), and sousveillance (``little watching'')~\cite{mann2002sousveillance, fletcher2011, negativesousveillance, michael2012sousveillance, bradwell2012security, bakir2010sousveillance, Freshwater2013Revisiting, ganascia2010generalized, Weston2009Embracing, danaher2014sousveillance}.

\begin{figure*}
     \centering
     \includegraphics[width=\textwidth]{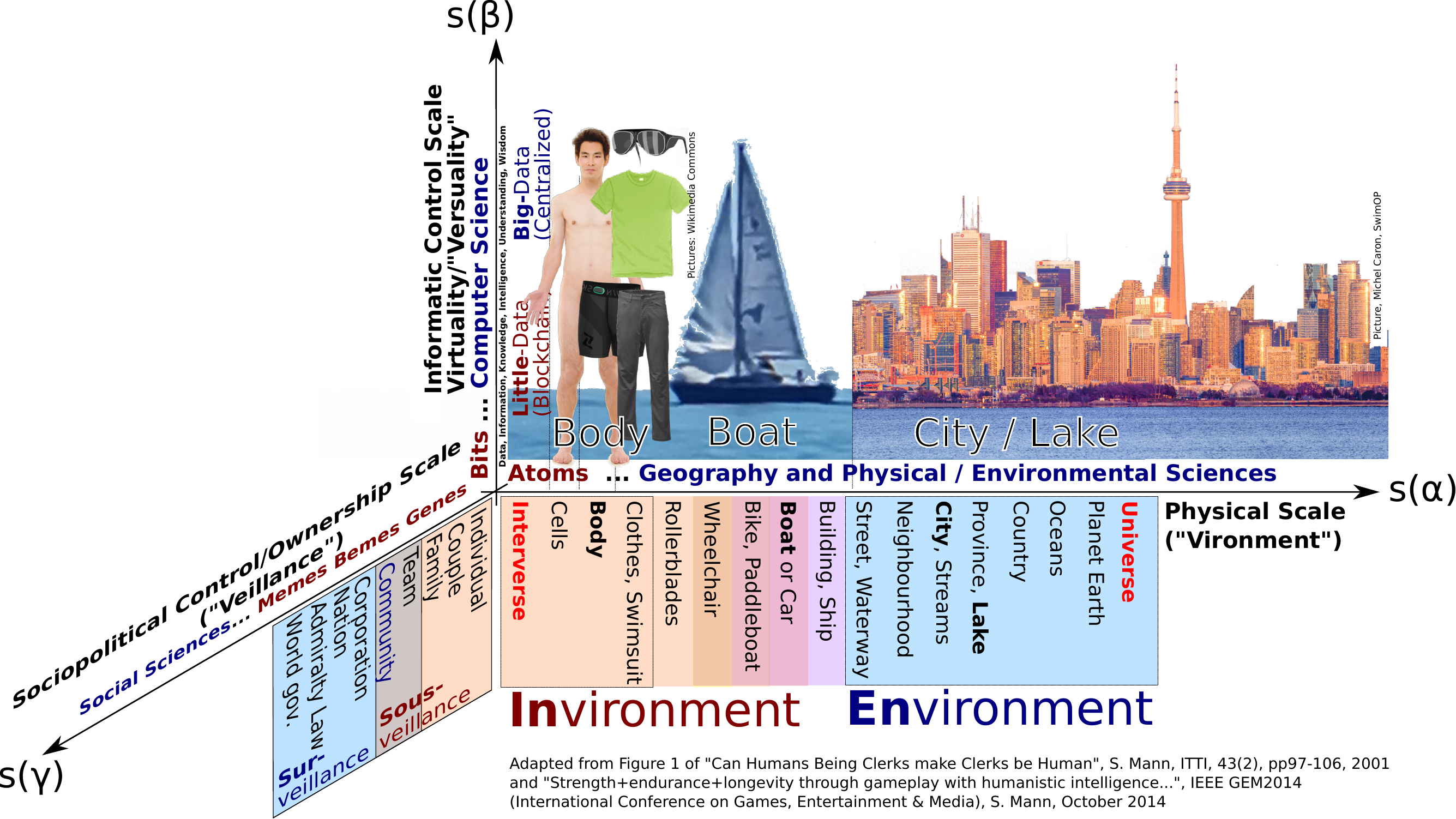}
     \caption{A three-dimensional taxonomy based on Physical Scale, $s(\alpha)$, Virtual Scale, $s(\beta)$, and Social Scale, $s(\gamma)$.}
     \label{fig:urbanscalespace}
 \end{figure*}

\begin{figure*}
     \centering
     \includegraphics[width=0.95\columnwidth]
     {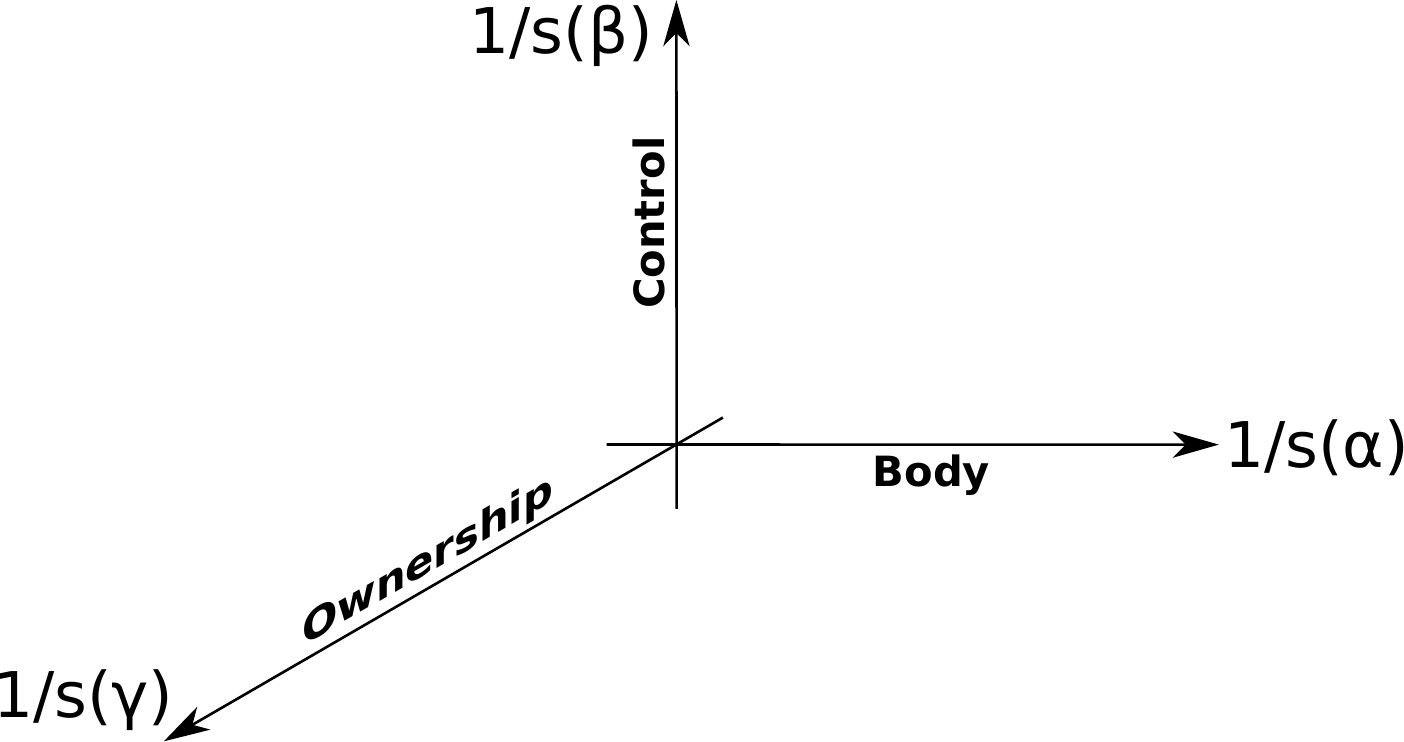}
     \includegraphics[width=\columnwidth]{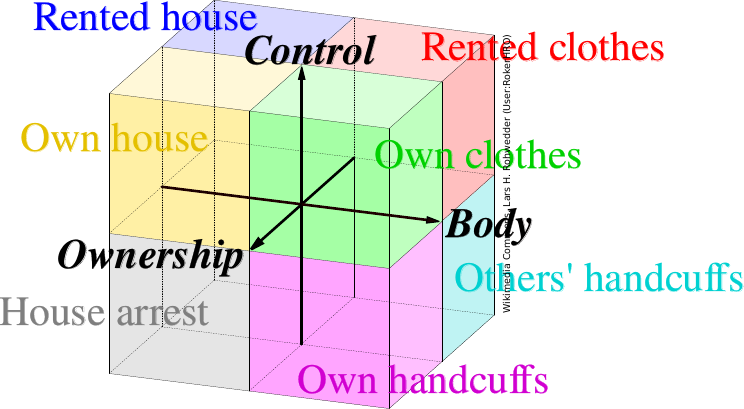}
     \caption{The three-dimensional taxonomy in terms of Body, Ownership, and Control.  Most of the time we own and control that which we wear (shoes, clothing, eyeglasses, etc) with a few exceptions like prison or work uniforms, handcuffs, etc.  Increasingly, though, we may find ourselves losing ownership or control of that which is close to or even inside our body.}
     \label{fig:boc}
 \end{figure*}

We now have a taxonomy of scale, based on $s(\alpha)$,
$s(\beta)$, and $s(\gamma)$, as illustrated in Fig~\ref{fig:urbanscalespace}
This taxonomy breaks the space up into 8 octants, e.g. wearable versus not wearable, little data (e.g. blockchain) versus big data (centralized), and sousveillance (self-sensing) versus surveillance.
Each of us has our own such space around our own body, as we interact with the 3D space around others.
Reversing the directions of each of these three axes gives us a new taxonomy based on
Body, Ownership, and Control, as shown in Fig~\ref{fig:boc}.
Again, each of us has our own set of coordinates that follow with us.
Generally we each have control over, and ownership of, technologies like shoes, clothing, and eyeglasses that we wear, versus lack of control or ownership of that which surrounds us (e.g. the cityscape).
Counterexamples are few but notable, e.g. handcuffs, which we can be wearing but not have control over.
Unravelling the complicated mess that ensues when the technologies that we wear are no longer controlled or owned by us, will be central to making XV into a technology in service of humanity!

\newpage
\section{Grand Challenges}
Finally we identify five grand challenges:
\begin{enumerate}
 \item {\bf BOC = Body, Ownership, and Control} are the central issues surrounding the ethical dilemmas created by body-borne (in-body, on-body, and close-to-body technologies) like ``wearables'' and ``implantables''\cite{mannieeecomputer}.
 The central objective we have at IEEE is ``Advancing technology for humanity'', and these issues are central to ethics, law, governance, freedom, democracy, privacy, security, digital identity, centralized versus distributed power and control, trust and trustworthiness.
 \item {\bf Standards} often evolve in a haphazard market-driven way.  We seek to create
 {\em market-driving} standards rather than {\em market-driven} standards.
 This is especially important of ``cyborg'' technologies that become part of us, in the sense of being wearable in our everyday lives.
 XV standards will be an important part of the IEEE Standards Association's activities in the coming years.
 \item {\bf Human factors + perceptual effects} are very important.  So far our taxonomy, for the most part, ignores these.  At present our taxonomy categorizes technology based on ``what it is'' rather than ``what it does''.  Much remains to be done to understand and design for perception. 
 Perception also becomes much more fluid and malleable than what our five senses directly produce.  Although we reference physical reality when describing sensors and systems on or near the body, in practice our perception may tunnel into distributed sensors across the planet in graceful ways that can effectively blur our sense of presence and scale~\cite{dublon2014sensor}.   ``Visualizers'' or perhaps rather ``sensory renderers'' working in the nascent new media inherent in this space won’t necessarily produce direct 1-1 sensory mappings, but will rather leverage perception in new ways by blending modalities, leveraging pre-cognitive human response~\cite{ananthabhotla2021cognitive}, and translating physically sensed and computationally inferred phenomena into media like music and animation~\cite{mayton2017networked, paradiso2018beyond}.
 \item {\bf Reliability} is of extreme importance to technologies that become part of us, or are closely integrated with our perception and understanding of the world.  Present-day XV technologies are challenged by everyday human activities as we run, dance, sweat, and swim, the electrical circuits need to be protected from water ingress, dust, dirt, and the rigours of day-to-day life.  Ruggedization remains a challenge for devices that must also be flexible and conform to the human body. 
 \item {\bf Storage, transmission, and data compression} are key challenges\cite{coughlin2011digital}, especially as many XV technologies are data intensive, and collect massive amounts of data often in a wearable or tetherless situation.  This will require 5g, 6g, 7g, 8g, 9g, 10g, etc., robust wireless networks.
 A typical XR glass might have a dozen or so high-resolution cameras plus numerous 3-dimensional depth-sensors.
 The idea of livelong sousveillance~\cite{mann2002sousveillance, fletcher2011, negativesousveillance, michael2012sousveillance, bradwell2012security, bakir2010sousveillance, Freshwater2013Revisiting, ganascia2010generalized, Weston2009Embracing, danaher2014sousveillance} is coming to fruition,
 both for personal safety and security\cite{grzonkowski2011security}
 as well as for health and wellbeing.
 Even of the storage is at least partially virtualized\cite{coughlin2010virtualization}, the communications bandwidth requirements will then also increase.
 Additionally, metadata is also an important consideration\cite{coughlin2010novel},
 and depending on how advanced the metadata is, storage and transmission requirements can further increase.
\end{enumerate}

\subsection{Future directions}
XV combines XR (eXtended Reality, perceived with senses or peripherals including neural-interfaces), XI (eXtended Intelligence, including HI = H. Int. = Humanistic Intelligence = human-in-the-loop Artificial Intelligence), XB (eXtended Being, including Digital Twin), XE (eXtended Economy), and XS (eXtended Society) into a vision that covers the generally agreed scope of metaverse while extending the impacts and emphasizing the implications on our consciousness and humanity.

Eventually we will reach a direct neural connection, but the same taxonomy will still persist, although direct neural connections are likely more easily able to augment than diminish one's own neural sensations.
There have been rapid advances in areas like auditory, ocular, and biomechatronic interfaces for people who are injured or sensorially impaired.  We argue that the fundamental XV taxonomy will still exist, with similar issues in suppressing or gracefully augmenting the information already feeding into the brain from our connected bodies and scaling experiences from full immersion to peripheral awareness.
Direct neural interfaces will need to work in a natural humanistic way to be effective, i.e. to be effective they will need to be affective~\cite{picard2000affective, pierce2021wearable}.

We are a long way from simulating an icewater swim in a virtual world,
and many challenges remain.  Perhaps the most difficult will be to earn trust, and solve the BOC-related issues surrounding ownership and control.

\section{Conclusion}
We have proposed XV as social, shared, collaborative XR, and provided a taxonomy in which to position the various existing realities (AR, VR, etc.) and verses (metaverse, universe, omniverse, etc.) within it.
We have also identified five grand challenges for XV:
(1) Standards; (2) Human factors and perceptual effects; (3) Reliability; (4) Storage and Transmission; and (5) Ethics and in particular, Body, Ownership, and Control, for which we have made a small adjustment to the proposed taxonomy to encompass.
Much work remains to be done, and we hope that many will join us to take on this important work!

\section{Acknowledgements}
The authors wish to thank Jon Paff, Cayden Pierce, Scott Williams, and the many other members of IEEE, IEEE SA, OpenXV, and SwimOP, for their useful suggestions and input.  We wish to also remember Micheal Hough, the landscape architect who created the beach where we do much of our research and teaching.  Perhaps we ought to name it ``Michael Hough Beach'' in his honour!

\bibliographystyle{IEEEtran}
\bibliography{references7}

\end{document}